\documentclass[]{article}
\usepackage{dcolumn}
\usepackage{epstopdf,epsfig}
\usepackage{graphicx,bm,bbm,amsmath,amssymb,amsgen,amsbsy,color}
\usepackage{hyphsubst,hyperref,lineno,fullpage}

\providecommand\bnabla{\bm{\nabla}}
\providecommand\bcdot{\bm{\cdot}}

\providecommand\der{\mathrm{d}}
\renewcommand{\vec}[1]{\ensuremath{\mathbf{#1}}} 
\newcommand{\gvec}[1]{\ensuremath{\mbox{\boldmath$ #1 $}}} 
\providecommand\bnabla{\gvec{\nabla}}

\newcommand{\vecd}[1]{\ensuremath{\overline{\mathbf{#1}}}}

\newcommand{\derv}[2]{\frac{\partial \overline{#1}}{\partial \overline{#2}}}

\newcommand{\aderv}[2]{\frac{\partial {#1}}{\partial {#2}}}

\newcommand{\degree}{\ensuremath{^\circ}}
\newcommand{\equa}[1]{Eq.~(\ref{#1})}
\newcommand{\fig}[1]{Fig.~\ref{fig#1}}
\newcommand{\tab}[1]{Table~\ref{tab#1}}
\newcommand{\equas}[1]{Eqs.~(\ref{#1})}
\newcommand{\equass}[2]{Eqs.~(\ref{#1})--(\ref{#2})}
\newcommand{\equasa}[2]{Eqs.~(\ref{#1}){ }and{ }(\ref{#2})}

\newcommand{\El}{\mathrm{El}}
\newcommand{\Ra}{\mathrm{R}}
\newcommand{\Pe}{\mathrm{Pe}}

\newcommand{\eqn}[2]{\begin{gather}
#1
\label{#2}
\end{gather}
}

\newcommand{\spll}[2]{\begin{gather}
\displaybreak[2]
\begin{gathered}
#1
\end{gathered}
\label{#2}
\end{gather}
}
\newcommand{\gat}[2]{\begin{subequations}\label{#2}\begin{gather}
#1
\end{gather}\end{subequations}
}

\usepackage{xcolor}

\title{Rayleigh--B\'enard instability of an Ellis fluid saturating a porous medium}

\author{%
\textsc{Michele Celli, Antonio Barletta and Pedro V. Brand\~ao} \\[1ex] 
\normalsize \href{mailto:michele.celli3@unibo.it}{michele.celli3@unibo.it} 
\normalsize \href{mailto:antonio.barletta@unibo.it}{antonio.barletta@unibo.it} 
\normalsize \href{mailto:pedro.vayssiere2@unibo.it}{pedro.vayssiere2@unibo.it} \\
\normalsize Alma Mater Studiorum Universit\`a di Bologna, Department of Industrial Engineering,\\
\normalsize Viale Risorgimento 2, 40136 Bologna, Italy 
}

\date{\today}
\begin{document}
\maketitle
\begin{abstract}
Unlike the power--law model, the Ellis model describes the apparent viscosity of a shear--thinning fluid with no singularity in the limit of a vanishingly small shear stress. In particular, this model matches the Newtonian behaviour when the shear stresses are very small. The emergence of the Rayleigh--B\'enard instability is studied when a  horizontal pressure gradient, yielding a basic throughflow, is prescribed in a horizontal porous layer. The threshold conditions for the linear instability of this system are obtained both analytically and numerically. In the case of a negligible flow rate, the onset of the instability occurs for the same parametric conditions reported in the literature for a Newtonian fluid saturating a porous medium.  On the other hand, when high flow rates are considered, a negligibly small temperature difference imposed across the horizontal boundaries is sufficient to trigger the convective instability.
\end{abstract}



\section{Introduction}

{The investigation of the threshold conditions for the onset of buoyancy--driven convection of non--Newtonian fluids is a research topic that displayed a significant development in the last decades \cite{shenoy1994,NiBe17,metivier2017origin,taleb2016analytical,khechiba2017effect,Gri2017,delenda2012}.}
Within this area of fluid dynamics, the shear--thinning fluids, also well--known as pseudoplastic fluids, are extremely common. Pseudoplastic fluids are important in different research areas. For instance, polymer solutions display shear--thinning behaviour. The same happens for some biological fluids like blood and a significant number of liquid foods {\cite{shenoy1994,van2015non,yoshida2019ultrasonic}.}

The viscosity of pseudoplastic fluids is often described by employing the Ostwald--De Weale (power--law) model. The drawback of this model is in its singular behaviour for negligibly small shear stresses. In fact, for this particular case, the power--law model predicts that pseudoplastic fluids display an infinite apparent viscosity \cite{bird1965}.  The Ellis model is employed to overcome this issue. This rheological model yields the Newtonian viscosity when the shear stresses applied to the shear--thinning fluid are extremely small \cite{bird1987}. 

The analysis presented in this paper is aimed to study the threshold conditions for the onset of buoyancy--driven convection in shear--thinning fluids saturating a porous medium. Since the stresses involved at onset of thermal instability may be negligibly small, the Ellis model will be employed. More precisely, the Rayleigh-B\'enard instability will be analysed when an Ellis fluid saturates a horizontal porous layer. Isothermal impermeable boundaries kept at different temperatures are envisaged providing a heating--from--below condition. In perspective, the results of this study are important as they can be suitable for an experimental validation by using, for instance, a Hele--Shaw cell system \cite{celli2017thermal}. In fact, the most unstable rolls for shear--thinning fluids were predicted to be transverse \cite{barl11,celli2018onset}, by employing a power--law model.

\section{Mathematical Modelling}
The height of the horizontal porous layer is $H$ and the boundaries of the layer are impermeable and isothermal such that a heating--from--below configuration is present. The lower boundary is held at temperature $T_0 + \Delta T$ (with $\Delta T>0$), while the upper boundary is held at temperature $T_0$, as displayed in \fig{1}. A basic throughflow is imposed by prescribing a horizontal pressure gradient. 
\begin{figure*}
  \centerline{\includegraphics[width=0.8\textwidth]{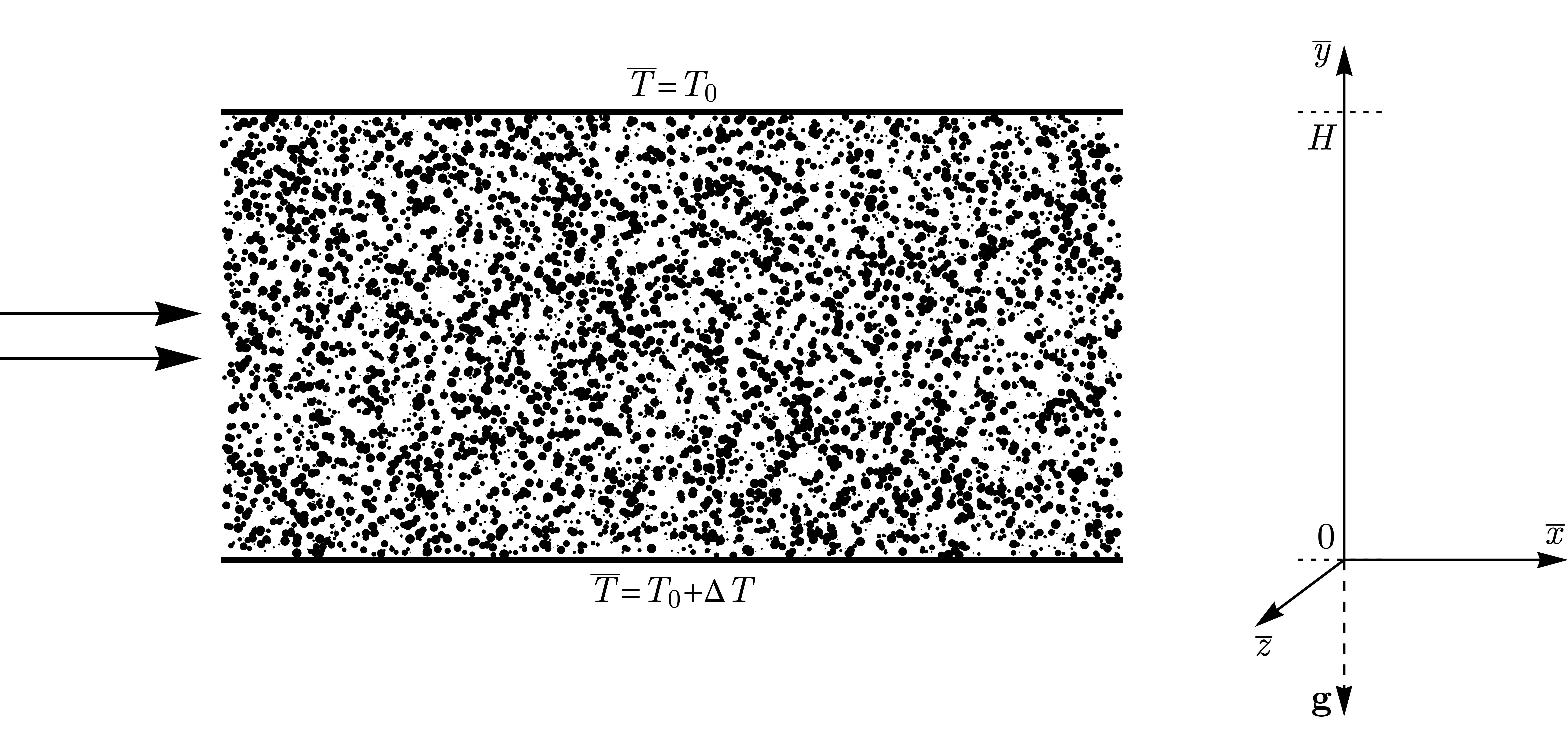}}
\caption{Sketch of the porous layer heated from below with horizontal throughflow}
\label{fig1}
\end{figure*}
\subsection{The Ellis model}\label{Ellis}
The rheological Ellis model defines the apparent viscosity $\eta$ of the non--Newtonian shear--thinning fluid as reported in \cite{savins1969}, namely
\spll{
\eta=\dfrac{\eta_0}{1+\left ( \dfrac{\tau}{\tau_0}\right)^{\frac{1-n}{n}} },
}{1} 
where $n$ is a positive parameter such that $0<n<1$, $\tau_0$ represents the value of $\tau$ at which the apparent viscosity drops by half its reference value $\eta_0$, $\tau$ is the scalar quantity
\spll{
\tau=\sqrt{\dfrac{\gvec \tau : \gvec \tau}{2}}.
}{0} 
Here $\gvec \tau$ is the shear stress tensor and {$\gvec \tau : \gvec \tau=\tau_{ij} \tau_{ij}$}, where the Einstein notation for the sum over repeated indices is implied. The behaviour of the viscosity ratio $\eta/\eta_0$ versus the shear stress ratio $\tau/\tau_0$ for different values of $n$ is reported in \fig{2} together with the behaviour of $\eta/\eta_0$ versus $n$  for different values of $\tau/\tau_0$. In the limiting cases of $\tau_0 \to 0$, $\tau_0 \to \infty$, $n\to 0$, $n \to 1$, Ellis model reduces to the Newtonian model as shown in \tab{2}.\\
On the other hand, when the fluid undergoes intense shear stresses, $\tau \gg \tau_0$, \equa{1} simplifies to
\spll{
\eta=\eta_0\left ( \dfrac{\tau}{\tau_0}\right)^{\frac{n-1}{n}} . 
}{2} 

\begin{table}[b]
\begin{center}
\caption{\label{tab2} Apparent viscosity for some limiting cases}
\begin{tabular}{c c c c c c}
 & $\tau_0\to 0$ & $\tau_0\to \infty$   & \multicolumn{2}{c}{$n\to 0$}  &   $n\to 1$  \\
\hline
 &  &    &   $\tau<\tau_0$ &   $\tau>\tau_0$ &   \\
\hline
$\eta$& 0& $\eta_0$& $\eta_0$&$0$&$\eta_0/2$\\
\end{tabular}
\end{center}
\end{table}

%
\subsubsection{Ellis model and power--law model}
The power--law fluid model prescribes that the apparent viscosity of the fluid 
be the following function of the 
shear stress:
\spll{
\eta =\chi^{\frac{1}{n}} \, \tau^{\frac{n-1}{n}},
}{4} 
where $\chi$ is the consistency factor and $n$ is the power--law index. The limiting case described in \equa{2} thus coincides with the power--law model \equa{4} if one defines $\chi=\eta_0^n \, \tau_0^{1-n}$.
\begin{figure*}
\centering
\includegraphics[width=1\textwidth]{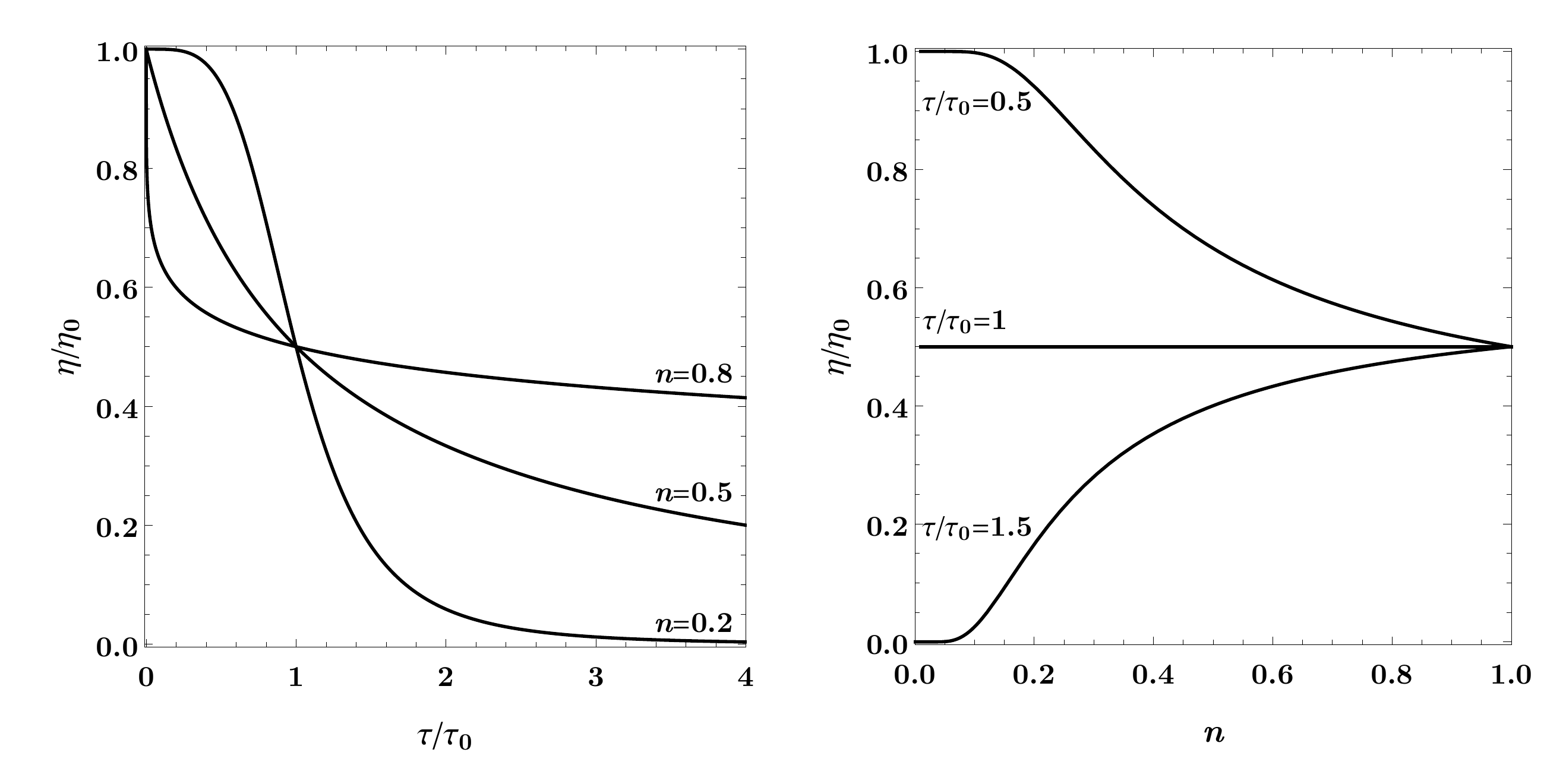}
\caption{Values of $\eta/\eta_0$ versus $\tau/\tau_0$ for different values of $n$, left--hand frame. Values of $\eta/\eta_0$ versus $n$ for different values of $\tau/\tau_0$, right--hand frame.}
\label{fig2}
\end{figure*}

\subsection{Modified Darcy's law for an Ellis fluid}
The  momentum balance equation for a Newtonian fluid saturating a porous medium is Darcy's law, namely
\spll{
 \vec u =\dfrac{K}{\eta} \,\vec f_d,
}{6} 
where $ \vec u$ is the filtration velocity vector of components $(u,v)$, $K$ is the permeability of the porous medium and $\vec f_d$ is the drag force defined as follows:
\spll{
\vec f_d=- \bnabla p - \rho_0 \, \vec g \, \beta \, ( T-T_0).
}{7} 
In \equa{7} the Oberbeck-Boussinesq approximation is invoked, $p$ is the pressure head, $\rho_0$ is the fluid density evaluated at the reference temperature $T_0$, $\vec g$  is the gravity acceleration vector and $\beta$ is the thermal expansion coefficient of the fluid. A modified Darcy's law that describes a porous medium saturated by an Ellis fluid has been proposed by \cite{Sad}, as well as by \cite{SadII}, namely
\spll{
 \vec{u} =\dfrac{K}{\eta_0} \left( 1 + A\, |\vec{f}_d|^{\frac{1-n}{n} } \right) \vec{f}_d,
}{8} 
where $A$ is a fluid property $[({\rm Pa/m})^{1-1/n}]$. In the limiting case of $A |\vec f_d|^{1/n-1} \ll 1$, that is when negligible drag forces are acting on the fluid, \equa{8} matches Darcy's law (\ref{6}). It is worth noting that at the onset of natural convection the intensity of the drag forces may be negligibly small.
\subsection{Governing equations}
The governing equations describing the problem here presented are
\spll{
\overline \bnabla \bcdot \vecd{u} = 0,\\
\frac{\eta_0}{K}\; \vecd{u} = \left( 1 + A\, |\vecd{f}_d|^{\frac{1-n}{n}} \right) \vecd{f}_d  ,\\
\sigma\, \derv{T}{t} + \vecd{u} \bcdot \overline \bnabla \overline T = \alpha \, \overline \nabla^2 \overline T,\\
\overline y=0 : \qquad \overline v = 0, \quad \overline T = T_0 + \Delta T,  \\
\overline y=H : \qquad \overline v = 0, \quad \overline T = T_0 ,
}{9}
where the bars over the quantities identify dimensional fields, coordinates and time, $\sigma$ is the ratio between the average volumetric heat capacity of the porous medium and the volumetric heat capacity of the fluid, and $\alpha$ is the average thermal diffusivity of the saturated porous medium. The drag force $\vecd f_d$ is given by \equa{7}.  
The following scaling allows us to express \equa{9} in a dimensionless formulation:
\spll{
\vec x=\frac{\vecd x}{H}, \quad  \vec u=\dfrac{H}{\alpha}\vecd u, \quad    p = \dfrac{K}{\eta_0 \, \alpha}\overline p ,\\
  t = \dfrac{\alpha}{\sigma H^2} \overline t, \quad T=\dfrac{\overline T-  T_0}{\Delta T }, 
}{10}
where $\vec x$ is the Cartesian position vector of components $(x,y,z)$. By substituting \equa{10} into \equas{9} one may write
\gat{
 \bnabla \bcdot \vec{u} = 0,\label{11a}\\
 \vec u= \left (1+\El \, |\vec f_d|^{\frac{1-n}{n}} \right) \vec f_d ,\label{11b}\\
\aderv{T}{t} + \vec{u} \bcdot  \bnabla  T =  \nabla^2  T,\\
 y=0 : \qquad  v = 0, \quad  T = 1, \label{11d} \\
 y=1 : \qquad  v = 0, \quad  T = 0 ,\label{11e}
}{11}
where
\spll{
\vec f_d= - \bnabla p +\Ra \, T\, \vec e_y.
}{fd}
The parameter $\El$ is the Darcy--Ellis number and the parameter $\Ra$ is the Darcy--Rayleigh number. They are defined as follows:
\eqn{
 \El=A\left(\dfrac{\alpha \, \eta_0}{H \, K}\right)^{\frac{1-n}{n}}, \qquad \Ra=\dfrac{\rho \, g \, \beta \, H \, K \, \Delta T}{\alpha \, \eta_0}.
}{12}
\subsection{Basic state}\label{basic}
The stationary solution of \equas{11} employed for the stability analysis is composed by a fully developed basic flow along the horizontal direction and a purely vertical constant temperature gradient. The horizontal flow is assumed to be generated by a prescribed {pressure gradient, which is independent of the $x$ and $z$ coordinates, such that}
\spll{
u_b =-\aderv{p_b}{x}\left (1+\El \, \left |\aderv{p_b}{x}\right |^{\frac{1-n}{n}} \right) , \quad v_b=0, \\ w_b=0,  \quad \aderv{p_b}{y}=\Ra \, T_b, \quad \aderv{p_b}{z}=0,\quad T_b = 1 - y ,
}{13}
where the subscript $b$ denote the basic state fields. It is not restrictive to assume that $\partial{p_b}/\partial {x}\leqslant 0$ so that $u_b \geqslant 0$. By taking the average value of the velocity profile, one obtains the definition of the P\'eclet number, namely
\spll{
\Pe=\int_0^1 u_b \, \der y  \qquad \longrightarrow \\
 \Pe=\left |\aderv{p_b}{x}\right|\left (1+\El \, \left |\aderv{p_b}{x}\right |^{\frac{1-n}{n}} \right) .
}{14}
For $\El \to 0$ with $|\partial{p_b}/\partial {x}|\ne 0$ one may simplify \equasa{11}{13} to obtain the basic state employed by the Prats problem \cite{PR66}. For $\El \to 0$ with $|\partial{p_b}/\partial {x}|=0$ \equasa{11}{13} yields the basic state employed by the Horton--Rogers--Lapwood problem \cite{HR45,L48}. 
{\subsection{Pressure--temperature formulation}
By employing \equa{11a} and by applying the divergence operator to \equa{11b}, we can express \equas{11} according to a pressure--temperature formulation,
\gat{
\bnabla \bcdot \left [\left (1+\El \, |\vec f_d|^{\frac{1-n}{n}} \right)  \vec f_d \right]=0 ,\\
\aderv{T}{t} +\left[ \left (1+\El \, |\vec f_d|^{\frac{1-n}{n}} \right) \vec f_d \right]   \bcdot  \bnabla  T =  \nabla^2  T,\\
 y=0 : \qquad  \aderv{p}{y} = R, \quad  T = 1,\label{15d}  \\
 y=1 : \qquad  \aderv{p}{y}  = 0, \quad  T = 0 , \label{15e}
}{15}
where the impermeability conditions in \equasa{11d}{11e} result into pressure conditions,  \equasa{15d}{15e}.}
\section{Linear stability analysis}
 The system~(\ref{15}) is perturbed by defining the pressure and temperature fields as composed by a basic state plus small--amplitude disturbances expressed in terms of normal modes, namely
\spll{
p = p_b + \varepsilon\, f(y)\, e^{\lambda \, t } e^{i \left( k_x \,  x +k_z \,  z - \omega \, t \right) }, \\
 T = T_b + \varepsilon\, h(y) \, e^{\lambda \, t } e^{i \left( k_x \,  x +k_z \,  z - \omega \, t \right) } .
}{16}
Here, $f$ and $h$ are, in general, complex functions, $\lambda$ is the growth rate, $\vec k=(k_x,0,k_z)$ is the wave vector, $\omega$ is the angular frequency. By assuming that the disturbance amplitude is small, $\varepsilon \ll 1$, we perform a linear stability analysis where we consider only terms $O(\varepsilon)$. The aim of the forthcoming investigation is finding the threshold for the onset of thermal convection. This threshold is obtained when the neutrally stable modes are considered. These modes are characterised by null growth rate. Thus, from now on, $\lambda$ is set equal to zero. By substituting \equa{16} into \equas{15}, and by employing
\spll{
\tilde f=(1+\tilde{\El})\, f, \; \; \; \;    \tilde \Ra=(1+\tilde{\El})\, \Ra,  \; \; \;\;  \tilde \omega=\omega -k \, \Pe, \\
 \tilde{\El}=\El  \left |\aderv{p_b}{x}\right |^{\frac{1-n}{n}}, \quad k_x=k\, \cos \phi , \quad k_z=k\, \sin \phi, \\
 \tilde n=\dfrac{ \tilde{\El}+n\, (\tilde{\El}+2) +\tilde{\El}\,  (1-n) \cos (2 \phi )}{2 \, n \,  (\tilde{\El}+1) },\\ 
}{sca}
 one obtains
\gat{
\tilde f''-\tilde n \, k^2\,  \tilde f-\tilde \Ra \, h'=0,\label{epa}\\
h''- \left(k^2- \tilde{\Ra}-i \, \tilde \omega \right)h-\tilde  f'=0,\label{epb}\\
 y=0,1 : \qquad  \tilde  f'= 0, \quad  h = 0,\label{epc}
}{ep}
where $\phi$ is the inclination angle between the wave vector and the $x$--axis. For $\phi=0$ the wave vector is parallel to the $x$--axis so that the rolls axes are perpendicular to the basic flow (transverse rolls). For $\phi=\pi/2$ the wave vector is parallel to the $z$--axis. In this case, the rolls axes are parallel to the basic flow (longitudinal rolls). In Appendix~\ref{appA} we prove analytically that $\tilde \omega=0$ and, hence, we conclude that the eigenvalue problem~(\ref{ep}) features real eigenfunctions and eigenvalues. It is worth noting that the P\'eclet number is not present, at least explicitly, in \equas{ep}. The definition of the rescaled angular frequency is a classical practice \cite{barletta2009darcy,barletta2009onset} for this kind of problems that follows the Prats choice \cite{PR66} of performing the stability analysis in the comoving reference frame.\\
On account of \equasa{14}{sca}, one may obtain $\Pe$ as a function of $ \El$, $\tilde \El$ and $n$, namely
\eqn{
 \Pe=\left (1+\tilde\El  \right) \left ( \dfrac{\tilde \El}{\El} \right)^{\frac{n}{1-n}} .
}{14a}
\begin{figure}
\centering
\includegraphics[width=0.45\textwidth]{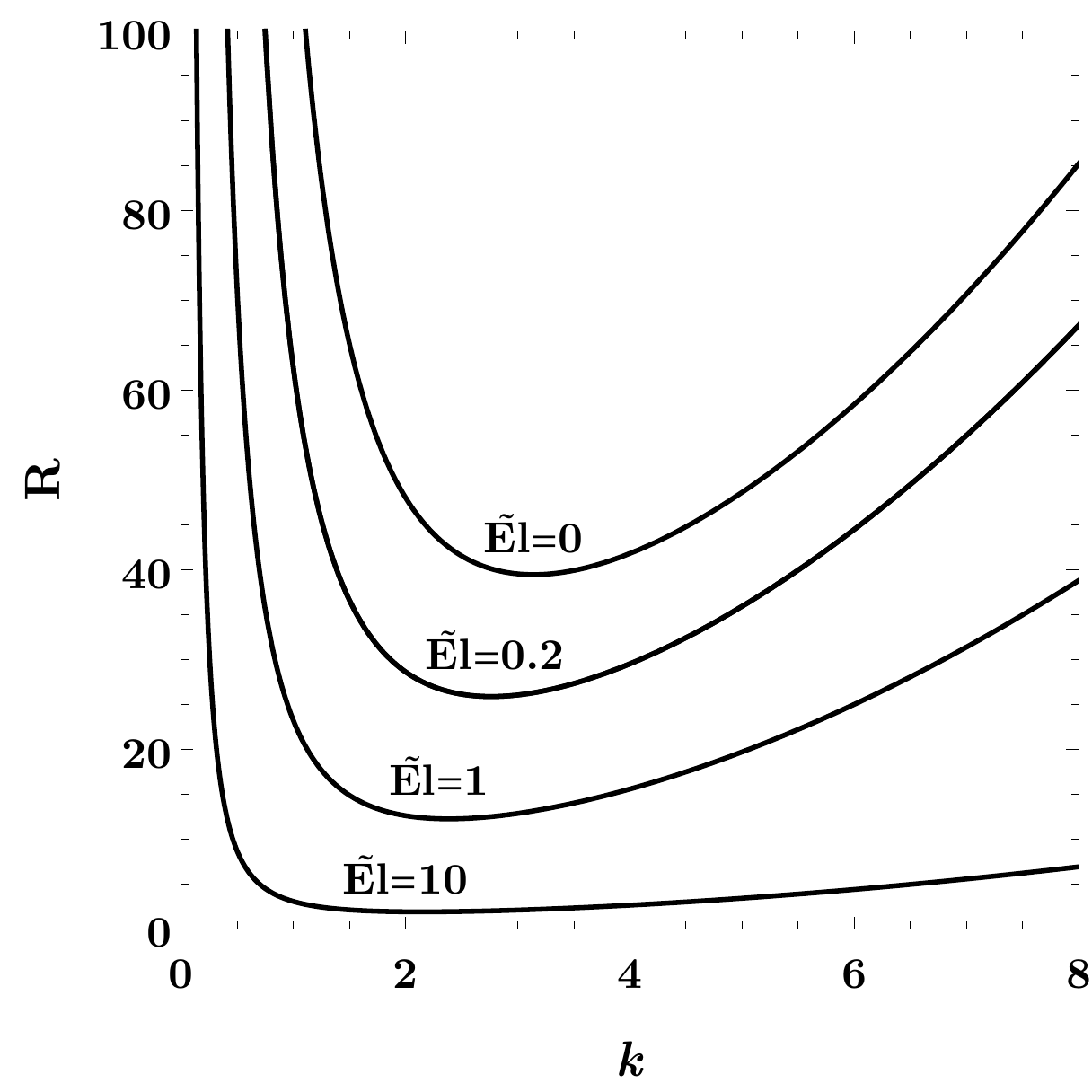}
\caption{Neutral stability curves $\Ra(k)$ for $n=0.2$ and different values of $\tilde{\El}$.}
\label{fig3}
\end{figure}
\section{Results}
The eigenvalue problem~(\ref{ep}) is solved both numerically and analytically. The numerical procedure, reported in Appendix~\ref{B}, is employed for comparison with the results obtained analytically. We assume that $\tilde f$ and $h$ are trigonometric functions satisfying the boundary conditions in \equa{epc}, namely
\spll{
\tilde f=\cos(\ell \, \pi \, y), \qquad  h=B_{\ell} \, \sin(\ell \, \pi  \, y),
}{23}
where $\ell$ is a positive integer, and $B_{\ell}$ is the constant. By employing \equass{sca}{23}, one obtains the dispersion relation
\spll{
 \Ra=\dfrac{k^2+\pi ^2 \, \ell^2}{(\tilde{\El}+1)}+\\\dfrac{2\, \pi^2 \,\ell^2\, n  \left(k^2+\pi ^2 \,\ell^2\right)}{k^2 [\tilde{\El}\, (1-n) \cos (2 \phi )+\tilde{\El} \,(n+1)+2 \, n]}.
}{24}
The most relevant parametric configuration for the stability analysis is the one characterised by the lowest values of $\Ra$. It is worth noting that, in order to minimise the value of $\Ra$, the integer and positive parameter $\ell$ must be minimum, \textit{i.e.} $\ell=1$. Moreover, by recalling that $0\leqslant \phi \leqslant \pi/2$, the minimum values of $\Ra$ are obtained for transverse rolls, $\phi=0$, since this angle minimises the contribution of the second term in the right--hand side of \equa{24}. Thus, at the onset of instability, \equa{24} can be simplified to
\spll{
 \Ra=\dfrac{\left(k^2+\pi ^2\right) \left[k^2 (\tilde{\El} +n)+n\, \pi ^2 \,  (\tilde{\El} +1) \right]}{k^2 \,(\tilde{\El} +1)  (\tilde{\El} +n)}.
}{25}
 Equation~(\ref{25}) allows one to draw the neutral stability curves presented in \fig{3}. This figure is obtained for the sample $n=0.2$ and different values of $\tilde{\El}$. \\
The absolute minimum of each neutral stability curve defines the parametric threshold for the onset of convective instability. The term ``critical values'' is employed to denote these threshold values of the governing parameters $R$ and $k$. In order to obtain the critical values, we calculate through \equa{25} the derivative of $\Ra$ with respect to $k$. Hence, the critical values are given by
\spll{
\Ra_c=\frac{\pi ^2 }{1+\tilde{\El}}\Bigg \{\left [\frac{n \,(1+\tilde{\El})}{n+\tilde{\El}}\right ]^{1/2}+1\Bigg \}^2, \\ k_c=\pi  \left [\dfrac{n\, (1+\tilde{\El})}{n+\tilde{\El}}\right]^{1/4}.
}{26}
The values of $\Ra_c$ and $k_c$ given by \equa{26} are reported versus $n$  in \fig{4}, for different values of $\tilde{\El}$. These critical values are shown to be monotonic decreasing functions of the parameter $\tilde{\El}$, while they are monotonic increasing functions of $n$. The four limiting cases $\tilde{\El} \to 0$, $\tilde{\El} \to \infty$, $n \to 0$, and $n\to 1$ deserve some particular attention.
\begin{figure*}
\centering
\includegraphics[width=1\textwidth]{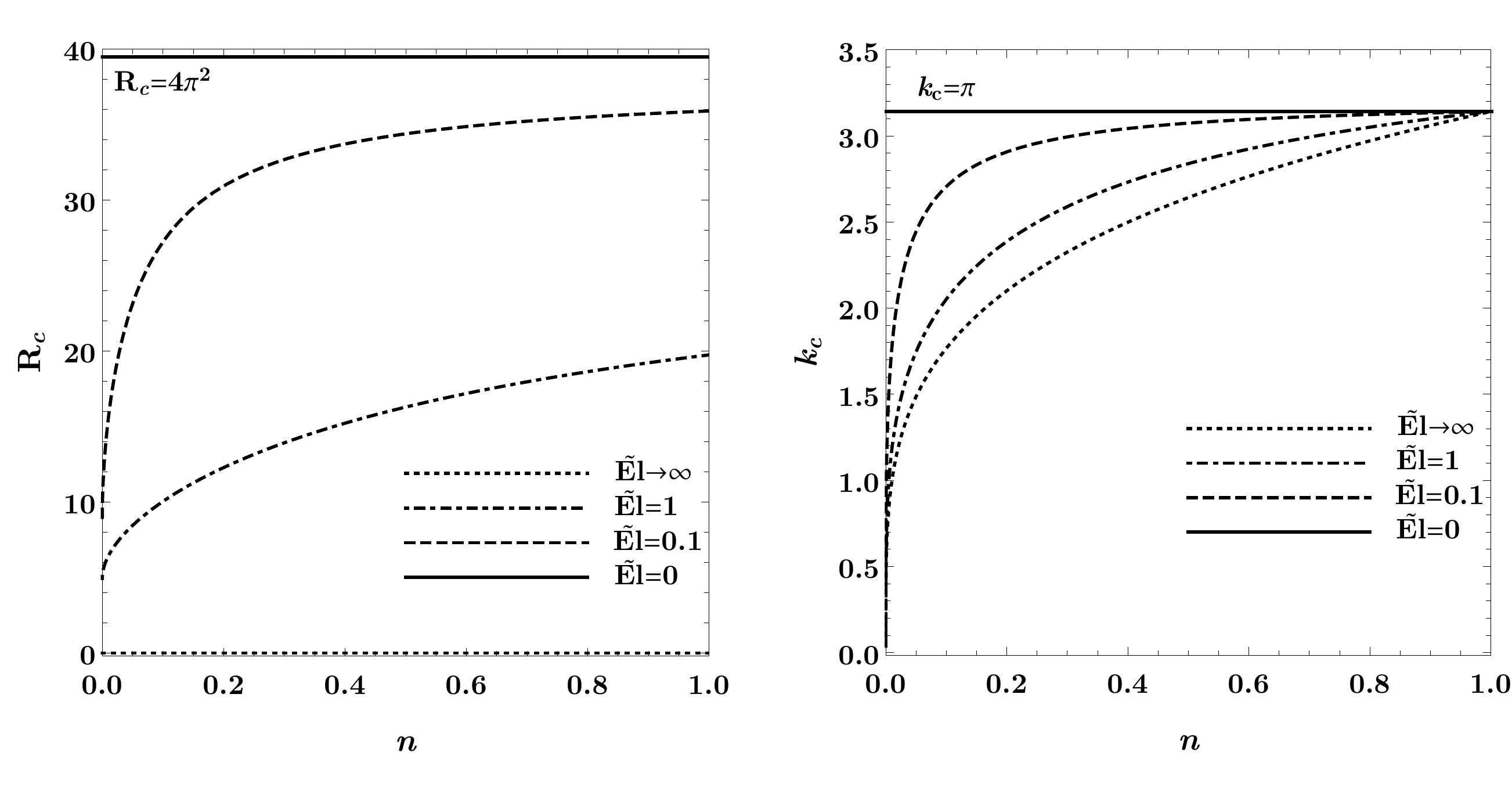}
\caption{Critical values of $\Ra$ and $k$ versus $n$ for different values of $\tilde{\El}$.}
\label{fig4}
\end{figure*}
\begin{figure*}
\centering
\includegraphics[width=1\textwidth]{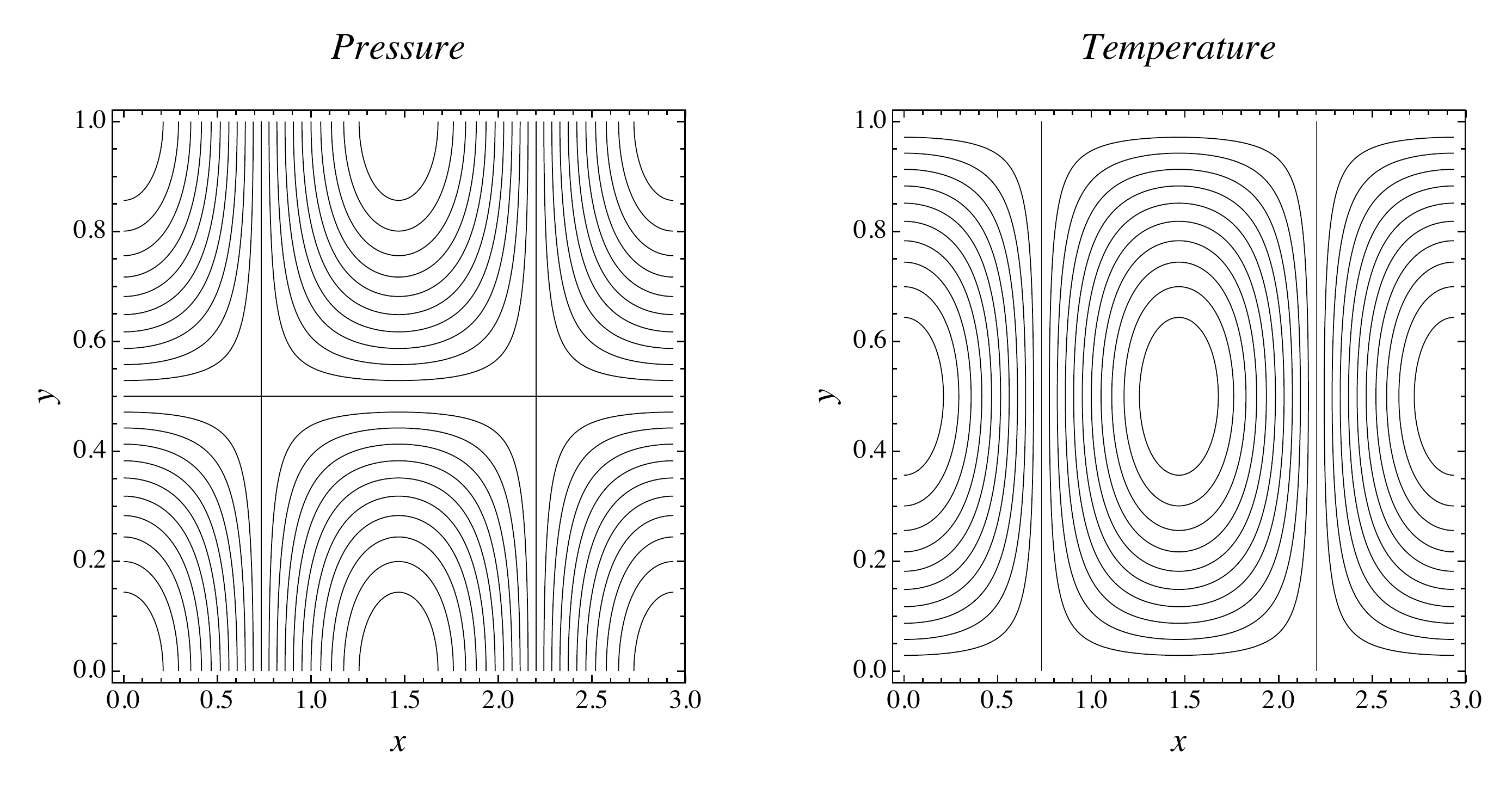}
\caption{{Lines $p(x,y,z,0)=constant$ (left frame) and lines $\theta(x,y,z,0)=constant$ (right frame). The figure is obtained for $n=0.2$ and $\tilde \El=1$.}}
\label{fig5}
\end{figure*}
\subsection{Limiting cases}
For a given value of $n$ such that $0< n < 1$, in the limiting case $\tilde{\El}  \to 0$, \equa{26} simplifies to
\spll{
\Ra_c=4 \, \pi^2 , \qquad k_c= \pi.
}{27}
The critical values given by \equa{27} coincide, as anticipated in Section~\ref{basic}, with those obtained by Prats \cite{PR66}, by Horton and Rogers \cite{HR45}, as well as by Lapwood \cite{L48}. \\
For  $0< n < 1$, in the limiting case $\tilde{\El} \to \infty$, \equa{26} simplifies to
\spll{
\Ra_c=0, \qquad k_c=\pi \,  n^{1/4}.
}{28}
{These results coincide with those reported in Barletta \& Nield \cite{BaNi} for the same limiting case}. For a finite non vanishing value of $\tilde{\El}$, in the limiting case $n  \to 0$ \equa{26}, simplifies to
\spll{
\Ra_c=\dfrac{\pi ^2}{\tilde{\El}+1} , \qquad k_c= 0.
}{29}
For a finite non vanishing  value of $\tilde{\El}$, in the limiting case $n  \to 1$ \equa{26}, simplifies to
\spll{
\Ra_c=\dfrac{4 \pi ^2}{\tilde{\El}+1}=\dfrac{4 \pi ^2}{{\El}+1} , \qquad k_c= \pi.
}{30}
Equations~{\ref{27}}~and~{\ref{29}} point out that the two limits $\tilde \El \to 0 $ and $n \to 0$ do not commute.
{\subsection{Shape of the disturbances}
Figure~\ref{fig5} displays the shape of the disturbances defined in \equa{16}. This figure is obtained by employing the critical wavenumber calculated for $n=0.2$ and $\tilde \El=1$ by means of \equa{26}. The lines defined by $p(x,y,z,0)=constant$ and the lines defined by $\theta(x,y,z,0)=constant$ are plotted for a single period $2 \pi / k$. Since the shape of the disturbances does not depend on the values of $n$ and $\tilde \El$, as one may infer from \equasa{16}{23}, only one case has been reported. Figure~\ref{fig5} refers to transverse rolls, $\phi=0$, and thus it is plotted on the plane $(x,y)$.}
\section{Conclusions}
The onset of convective instability inside a horizontal porous layer saturated by a non-Newtonian fluid has been investigated. The fluid is shear--thinning and its apparent viscosity is defined by the Ellis model. The layer is heated from below and a basic horizontal pressure gradient is assumed. A linear stability analysis has been performed by means of the normal mode method. The governing parameters are the Darcy--Rayleigh number, $\Ra$, the modified Darcy--Ellis number, $\tilde \El$, and the Ellis power--law index, $n$. The modified Darcy--Ellis number is a function of the P\'eclet number associated with the basic flow rate, of the Ellis number and of the Ellis power--law index. The main conclusions drawn from the stability analysis are the following:
\begin{itemize}
\item The critical values of the governing parameters can be expressed analytically as functions of $n$ and $\tilde \El$.
\item The most unstable rolls are transverse, having their axes perpendicular to the direction of the basic throughflow.
\item The angular frequency of the transverse rolls is equal to the product between the wavenumber and the P\'eclet number. Such rolls are non--travelling in the reference frame comoving with the basic throughflow.
\item For $\tilde \El\to 0$, the critical value of the Darcy--Rayleigh number tends to $4 \pi^2$ while the wavenumber approaches $\pi$. This limiting case identifies those configurations where the basic pressure gradient is absent and/or the fluid is Newtonian. The critical values of the governing parameters match those found in the literature for  either the Prats problem or the Horton--Rogers--Lapwood problem.
\item For $\tilde \El\to \infty$, the critical value of the Darcy--Rayleigh number tends to zero and the wavenumber tends to $\pi \,  n^{1/4}$. This limiting case identifies those configurations where the basic pressure gradient is extremely intense and/or the fluid is strongly shear--thinning. {In other words, for this parametric configurations, a fluid characterized by an extremely low apparent viscosity is considered and thus a negligibly small temperature gap between the horizontal boundaries is sufficient to trigger the onset of convection.}
\item The parameters $\tilde{\El}$ and $n$ play different roles: $\tilde{\El}$ has a stabilising effect on the basic state while $n$ has a destabilising effect.
\end{itemize}
{We finally point out that our study has been based on the Ellis model for the fluid rheology in order to encompass the singular behaviour of the simpler power--law model. In particular, as pointed out in Barletta \& Nield \cite{BaNi}, the use of the power--law model leads to the prediction of an either zero or infinite critical value of the Darcy--Rayleigh number when the flow rate in the basic state is zero. On the other hand, when the basic flow rate tends to zero, the use of the Ellis model leads to a non--singular behaviour where the same critical value of the Darcy--Rayleigh number as predicted for the case of a Newtonian fluid, namely $4\,\pi^2$, is attained.}
\section*{Aknowledgment}
This study was financed in part by the Coordena\c c\~ao de Aperfei\c coamento de Pessoal de N\'ivel Superior - Brazil (CAPES) - Grant n$\degree$ 88881.174085/2018--01.\\
 Financial support was also provided by Ministero dell'Istruzione, dell'Universit\`a e della Ricerca (Italy) -- Grant n$\degree$ PRIN2017F7KZWS.
\appendix
\begin{table*}
  \caption{  \label{tab1} Critical values of $\Ra$ and $k$ for $n=0.2$, $\phi=0$, and different values of $\tilde{\El}$. The subscript $a$ identifies those solutions obtained analytically while subscript $n$ identifies those solutions obtained numerically.}
 \begin{center}
  \begin{tabular}{lcccc}
      $\tilde{\El}$  & $k_{c,a}$ & $k_{c,n}$   &   $\Ra_{c,a}$ &   $\Ra_{c,n}$  \\
\hline
0.01& 3.11123554149690& 3.11123554149643& 38.3394030316440& 38.3394030316322\\
0.1& 2.90720213712325&2.90720213712280 &30.9190571158268&30.9190571158173\\
1& 2.38709420797841& 2.38709420797804&12.2779550251570&12.2779550251532\\
10& 2.14094494759125& 2.14094494759095&1.92414846749808&1.92414846749755\\
100& 2.10509056463515& 2.10509056463491&0.205169284151348&0.205169284151301\\
  \end{tabular}
  \end{center}
\end{table*}
\section{Proof that $\tilde \omega=0$}\label{appA}
One can multiply \equa{epa} by $\tilde f^*$, that is the complex conjugate of the eigenfunction $\tilde f$, and integrate by parts over the domain $y \in (0,1)$ to obtain
\spll{
\int_0^1 |\tilde f'|^2 \, \der y+k^2\, \tilde n \, \int_0^1|\tilde f|^2 \, \der y+\tilde \Ra\int_0^1 h' \tilde f^* \, \der y=0.
}{A3}
From \equa{A3}, one may conclude that the last integral on the left hand side is real. By taking the complex conjugate of this integral and, on integrating it by parts, one concludes that
\spll{
\int_0^1   \tilde f' h^*  \, \der y 
}{A4}
is real. 
This result will be invoked later on. One can now multiply \equa{epb} by $h^*$, that is the complex conjugate of the eigenfunction $h$, and integrate by parts over the domain $y \in (0,1)$ to obtain
\spll{
\int_0^1 |h'|^2 \, \der y+\left(k^2-\tilde {\Ra}-i \, \tilde \omega \right) \int_0^1 |h|^2 \, \der y+\\ 
\int_0^1 \tilde f' h^* \, \der y=0.
}{A6}
By employing \equa{A4}, one may infer that the imaginary part of \equa{A6} is
\spll{
\tilde \omega \int_0^1 |h|^2 \, \der y=0.
}{A7}
Equation~(\ref{A7}) implies either $\tilde \omega=0$ or $h=\tilde f=0$. Since the trivial solution $h=\tilde f=0$ is not acceptable, one may conclude that $\tilde \omega=0$.
\section{Numerical method}\label{B}
The numerical method employed to solve the stability eigenvalue problem is the shooting method. The first step consists in defining (and solving) the initial value problem obtained from \equa{ep} simplified as a consequence of the results reported in Appendix~\ref{appA}, namely
\spll{
\tilde f''-\tilde n \, k^2\,  \tilde f-\tilde \Ra h'=0,\\
h''- \left(k^2- \tilde{\Ra} \right)h-\tilde  f'=0,\\
\tilde f(0)=1, \quad \tilde f'(0)= 0, \quad  h(0) = 0,\quad h'(0)=\xi.
}{21}
Here, the condition $\tilde f(0)=1$ can be imposed because the governing equations in Eqs.~(\ref{21}) are homogeneous, while $\xi$ is an unknown real parameter. The problem~(\ref{21}) is solved numerically by means of the Runge--Kutta method. The obtained eigenfunctions $\tilde f$ and $h$ depend on four governing parameters, $(k,\tilde n, \tilde \Ra,\xi)$. \\
The second step of the shooting method is based on the target conditions
\spll{
\tilde f'(1)= 0, \quad  h(1) = 0 .
}{22}
Such conditions serve to obtain numerically, by employing a root--finding algorithm, two out of the four governing parameters $(k,\tilde n,\tilde \Ra,\xi)$. Thus, for every given $\tilde n$, one obtains the neutral stability curve $\tilde \Ra(k)$. \\
The critical values are obtained by solving the initial value problem given by \equa{21} and the derivative with respect to $k$ of \equa{21}. The conditions employed in the root--finding algorithm are the two conditions given by \equa{22} together with their derivatives with respect to $k$.\\
 A comparison between the results obtained analytically and those obtained numerically is reported in \tab{1}. The critical values of the wavenumber $k$ and the critical values of $\Ra$, both evaluated for $n=0.2$, $\phi=0$ and different values of $\tilde{\El}$, are provided in this table. The subscript $a$ refers to the data obtained analytically, while the subscript $n$ is relative to the numerical data. The results obtained by employing these two different approaches coincide within 12 significant figures.


\end{document}